\documentstyle[psfig,epsfig,graphicx,tikz]{mn}

\usetikzlibrary{shapes.geometric, arrows}

\newcommand{\be}{\begin{equation}}
\newcommand{\ee}{\end{equation}}
\newcommand{\ba}{\begin{eqnarray}}
\newcommand{\ea}{\end{eqnarray}}

\newcommand{\nn}{\nonumber \\}

\newcommand{\I}{\mbox{\boldmath $I$}}

\newcommand{\n}{\mbox{\boldmath $n$}}

\newcommand{\bell}{\mbox{\boldmath $\ell$}}

\newcommand{\mnras}{MNRAS}
\newcommand{\prd}{Phys. Rev. D}
\newcommand{\apjl}{Astro. Phys. Journal Letters}
\newcommand{\apj}{Astro. Phys. Journal}
\newcommand{\aap}{AAP}

\def\gs{\mathrel{\raise1.16pt\hbox{$>$}\kern-7.0pt %
\lower3.06pt\hbox{{$\scriptstyle \sim$}}}}         %
\def\ls{\mathrel{\raise1.16pt\hbox{$<$}\kern-7.0pt %
\lower3.06pt\hbox{{$\scriptstyle \sim$}}}}         %

\usetikzlibrary{shapes,arrows}

\title[The Limits of Cosmic Shear]{The Limits of Cosmic Shear}

\author[Kitching, Alsing, Heavens, Jimenez, McEwen, Verde]
       {Thomas D. Kitching$^1$\thanks{t.kitching@ucl.ac.uk}, Justin Alsing$^{2,3}$, Alan F. Heavens$^{2}$, Raul Jimenez$^{4,5}$, 
         \newauthor
         Jason D. McEwen$^1$, Licia Verde$^{4,5}$\\
         $^1$Mullard Space Science Laboratory, University College London, Holmbury St Mary, Dorking, Surrey RH5 6NT, UK\\
         $^2$ICIC, Astrophysics, Imperial College, Blackett Laboratory, Prince Consort Road, London SW7 2AZ, UK\\
         $^3$Center for Computational Astrophysics, 160 5th Ave, New York, NY 10010, USA\\
         $^3$ICC, University of Barcelona, IEEC-UB, Marti Franques, 1, E08028 Barcelona, Spain\\
         $^4$ICREA, Pg. Lluis Companys 23, 08010 Barcelona, Spain}

\date{}

\pagerange{\pageref{firstpage}--\pageref{lastpage}}

\pubyear{2015}

\begin{document}

\maketitle

\label{firstpage}

\begin{abstract}
In this paper we discuss the commonly-used limiting cases, or approximations, for two-point cosmic shear statistics. 
We discuss the most prominent assumptions in this statistic: 
the flat-sky (small angle limit), the Limber 
(Bessel-to-delta function limit) and the Hankel transform (large $\ell$-mode limit) approximations; that   
the vast majority of cosmic shear results to date have used simultaneously. 
We find that the combined effect of these approximations can suppress 
power by $\gs 1\%$ on scales of $\ell\ls 40$. 
A fully non-approximated cosmic shear study should use 
a spherical-sky, non-Limber-approximated power spectrum analysis; and 
a transform involving Wigner small-d matrices in place of the Hankel transform. 
These effects, unaccounted for, would constitute at least $11\%$ of the total 
budget for systematic effects for a power spectrum analysis of a Euclid-like experiment; 
but they are unnecessary. 
\end{abstract}


\section{Introduction}
\label{Introduction}
Weak lensing is the phenomenon whereby the images of distant galaxies are distorted by the effect of 
gravitational potentials caused by matter perturbations along the line-of-sight. This gravitational lensing 
effect induces a small change in the ellipticity\footnote{Third flattening, or third eccentricity.} of a 
galaxy's image known as shear. The shear caused by the large-scale structure of the Universe is known as `cosmic shear'. The 
mean of the complex cosmic shear field is zero but its 2-point correlation function or power spectrum contains 
cosmological information; 
cosmic shear is also used as a synonym for this statistic. This statistic is a particularly 
sensitive probe of dark energy because it measures the power spectrum of matter overdensity perturbations across 
large portions of the expansion history of the Universe. Because of this there are several on-going wide-field 
surveys that attempt to measure this effect, for example CFHTLenS (Heymans et al., 2012), DES (The DES Collaboration et al., 2015), 
DLS (Jee et al., 2015), KiDS (Kuijken et al., 2015), and HyperSuprimeCam; as well as 
several more planned surveys, for example \emph{Euclid}\footnote{{\tt http://euclid-ec.org}} (Laureijs et al., 2011), LSST 
(Tyson et al., 2003), and \emph{WFIRST} (National Research Council, 2010), that have the measurement of this statistic 
as one of their primary science goals. 

In practice there are several ways in 
which the cosmic shear 2-point statistic can be computed 
that can be broadly categorised into real/configuration-space measurements as a function of celestial angle 
(e.g., shear correlation functions), and angular spherical-harmonic/Fourier-space measurements (e.g.,  power spectra). 
Furthermore these statistics can be computed in a series of redshift bins, to capture 
the geometry of the three-dimensional shear field, an approach known as `tomography'; or a 
spherical-Bessel/Fourier-space measurement in 
the radial direction known as `three-dimensional' cosmic shear (Heavens, 2003, Castro et al., 2005, Kitching et al. 2007).

In this paper we present each of the primary approximations in cosmic shear 
statistics and explicitly link all of the currently used statistics together into a general schema. In doing so we also present 
a general three-dimensional spherical-radial statistic which is the redshift-space equivalent of a spherical-Bessel analysis. 
We discuss various approximations and a data compression, namely: flat-sky,  Limber, tomography and Hankel transformations. 
The flat-sky 
assumption projects onto a locally flat tangent plane on the sky. The tomographic 
data compression, presented in Hu (1999), is a lossy binning of the cosmic shear signal into several 
redshift bins and is an approach used by 
\emph{all} cosmic shear studies (see Kilbinger, 2015 for a review) 
except those that use a spherical-Bessel representation (e.g. Kitching et al. 2014), 
for both theoretical studies and data analysis. In Kitching, Heavens, Miller (2011) and 
Kitching et al. (2014) it was shown how to derive the tomographic 
case from a more general spherical-Bessel representation of the 
shear field. The Limber approximation links angular and radial wavenumbers together via a comoving distance relation. This was 
first discussed in Kaiser (1998) in the context of cosmic shear and has been investigated in 
Kitching, Heavens, Miller (2011) in cosmic shear studies, 
but in the majority of theoretical studies and data analyses it is an assumption. 
There is a particularly clear illustration of 
the Limber approximation in LoVerde \& Afshordi (2008) that we discuss in this paper. 

Most of the approximations we investigate are used simultaneously and in combination. Notably all the primary cosmological 
results from all of the wide-field surveys use a flat-sky, tomographic, Limber-approximated correlation 
function analysis, e.g.,
Heymans et al. (2013) for CFHTLenS; The DES Collaboration et al., (2015) for 
DES; Jee et al. (2015), and Hildebrandt et al. (2017) for KiDS. 
Notable exceptions include Pen et al. (2002), Brown et al. (2003), Heymans et al. (2005),
K{\"o}hlinger et al. (2016), Alsing et al. (2017), and the {\tt PolSpice} 
measurements in The DES Collaboration et al., (2015), 
all whom used Fourier-space measurements in angle, with the flat-sky, tomographic and Limber approximations. 
In Kitching et al. (2007) and 
Kitching et al. (2014) a flat-sky spherical-Bessel analysis was used without the tomographic or Limber approximations. 

This paper is presented in the following manner. 
In Section \ref{Method} we review the cosmic shear formalism starting with the spherical-Bessel representation and then present 
the spherical-radial and correlation function representations. In Section \ref{Cosmic Shear Approximations} we 
discuss the flat-sky, tomographic and Limber approximations and present a general 
schema for linking these statistics and approximations. 
We discuss the implications of this for current results in 
Section \ref{Discussion}. We discuss conclusions in Section \ref{Conclusion}.

\section{Cosmic Shear Methods}
\label{Method}
We begin by introducing several versions of the two-point cosmic shear statistic that treat the 
data, and represent the underlying three-dimensional shear field, in different ways. 
The first of these is the spherical-Bessel representation that has been described in detail in 
Heavens (2003); Castro, Heavens, Kitching (2005); Heavens, Kitching, Taylor, (2006); 
Kitching (2007); Kitching, Taylor, Heavens, (2008); Kitching, Heavens, Miller (2011);
Kitching et al. (2007); Kitching et al. (2014),  
the second is the presentation of a spherical-radial representation of which the commonly used 
tomographic statistic (Hu, 1999) is a simple approximation. We then discuss real/configuration-space representations. 

\subsection{The Spherical-Bessel Representation}
The cosmic shear field has spin-weight 2, and we can perform a spherical-Bessel transform
to obtain 
\be
\label{a}
\gamma^m_{\ell}(k)=\left(\frac{2}{\pi}\right)^{1/2}\sum_g \gamma_g(r_g,\theta_g) j_{\ell}(kr_g)_2Y^m_{\ell}(\theta_g)
\ee
where the sum is over all galaxies $g$ at three-dimensional comoving coordinates 
$(r_g$, $\theta_g)$, $k$ is a radial wavenumber and $\ell$ is an angular wavenumber. The 
$j(kr_g)$ are spherical Bessel functions. The $_2Y^m_{\ell}(\theta_g)$ are spin-weight 2 spherical harmonics. 
Such a sum can be used to construct the data vector for a spherical-Bessel analysis of 
weak lensing data, which is then compared with the following theoretical covariance, as described in Kitching et al. (2014). 
When applying this sum to data these transformed coefficients can be manipulated to extract the pure E and B-mode signals 
(where cosmic shear is only expected to produce an E-mode signal), and remove 
any multiplicative measurement biases (where the measured $\gamma_g$ is related to the 
true $\gamma^T_g$ via some linear relation $\gamma_g=(1+m)\gamma^T$, where $m$ is an estimated bias parameter) 
as described in Kitching et al. (2014). The sum over galaxies is an estimator for a continuous integral over angle and radius, 
where there is an additional shot-noise contribution to the covariance, due to having a finite number of galaxies at 
discrete points (see Heavens, 2003). The factor $(2/\pi)^{1/2}$ is a convention that is consistent with 
Heavens et al. (2006; equation 2). 

The mean of equation (\ref{a}) is zero, but the covariance of the transform coefficients is non-zero. Assuming 
isotropy the covariance of the harmonic coefficients -- known as the power spectrum -- can be written as 
\be
\langle \gamma^m_{\ell}(k)\gamma^{m'*}_{\ell'}(k')\rangle=C^{SB}_{\ell}(k,k')\delta_{\ell\ell'}\delta_{mm'}.
\ee
Using the notation of Kitching, Heavens, Miller (2011), we can write down the theoretical expectation value 
of the power spectrum for given a cosmology 
\be
\label{sb}
C^{SB}_{\ell}(k, k')=|D_{\ell}|^2{\mathcal A}^2\left(\frac{2}{\pi}\right)\int \frac{{\rm d}\tilde k}{\tilde k^2} G^{SB}_{\ell}(k,\tilde k)G^{SB}_{\ell}(k',\tilde k), 
\ee
where the pre-factor ${\mathcal A} = 3\Omega_{\rm M} H_0^2/(2 c^2)$ (where $H_0$ is the 
current value of the Hubble parameter, 
$\Omega_{\rm M}$ is the ratio of the total matter density to the critical density, and $c$ is the 
speed of light in a vacuum). The variable $|D_{\ell}| = \sqrt{(\ell+2)!/(\ell-2)!}$ in the spherical case (see 
Castro et al., 2005; and Leistedt et al., 2015).  The temptation in the flat-sky case is to 
approximate $|D_{\ell}|=\ell^2$, but this is an approximation. 
The $G$ matrix is given by 
\ba
G^{SB}_{\ell}(k,\tilde k)&=&\int {\rm d}z_p j_{\ell}(k r(z_p)) n(z_p)\nn 
&\times&\int {\rm d}z' p(z'|z_p) U_{\ell}(r[z'],\tilde k), 
\ea
where $n(z_p){\rm d}z_p$ is the number of galaxies in a spherical shell 
of radius $z_p$ and thickness ${\rm d}z_p$, and $p(z'|z_p)$ is the probability of
a galaxy with photometric redshift $z_p$ having a true redshift $z'$. The $U$ matrix is given by 
\be
\label{U}
U_{\ell}(r[z],k)=\int_0^{r[z]} {\rm d}r' \frac{F_K(r,r')}{a(r')} j_{\ell}(kr')P^{1/2}(k,r'), 
\ee
where $P(k,r[z])$ is the matter power spectrum at comoving 
distance $r[z]$ and radial wavenumber $k$. The comoving distance $r$ is used to express the time-dependence of 
the power spectrum; we could equally use $t$ as a label, or $r(t)$.
$F_K = S_K(r - r')/S_K(r)/S_K(r')$ is the `lensing kernel' where $S_K(r) = \sinh(r)$, $r$, $\sin(r)$ 
for cosmologies with spatial curvature $K = -1$, $0$, $1$, and $a(r)$ is
the dimensionless scale factor at the cosmic time related to the look-back time at comoving distance $r$.  
Note that already we have made an approximation, in that the statistics strictly depend on unequal-time correlators (Kitching \& Heavens 2016), but we will not discuss this point further here.
 
\subsection{The Spherical-Radial Representation}
A different way to represent the three-dimensional shear field is to make a Fourier-like decomposition in 
angular wavenumber but \emph{not} in the radial direction. This decomposition is the following
\be
\label{b}
\gamma^m_{\ell}(z)=\left(\frac{2}{\pi}\right)^{1/2}\sum_{g \in z} \gamma_g(r_g,\theta_g) _2Y^m_{\ell}(\theta_g)
\ee
which is still a three-dimensional representation of the data, except that it excludes the radial Bessel transform. 
The sum in this case is over all galaxies that have a redshift $z$. 
We refer to this as the `spherical-radial' transform (as opposed to a spherical-Bessel transform). 

Again the mean of this representation is zero, but the covariance is non-zero. 
Using the notation above, we can write down the theoretical expectation value
of the power spectrum given a cosmology
\be
\label{sr}
C^{SR}_{\ell}(z, z')=|D_{\ell}|^2{\mathcal A}^2\left(\frac{2}{\pi}\right)\int \frac{{\rm d}k}{k^2} G^{SR}_{\ell}(z,k)G^{SR}_{\ell}(z',k),
\ee
where in this case the $G^{SR}$ matrix is given by
\ba
\label{Gsr}
G^{SR}_{\ell}(z,k)&=&\int {\rm d}z_p W^{SR}(z,z_p) n(z_p)\nn 
&\times&\int {\rm d}z' p(z'|z_p) U_{\ell}(r[z'],k),
\ea
where $W(z,z_p)$ is a redshift-dependent weight function that defines the `bin-width' in redshift over which the 
statistic is defined for redshift $z$. The $U$ matrices are the same as in equation (\ref{U}). 
In the case that $W^{SR}(z,z_p)=\delta^D(z-z_p)$ 
this covariance is still a complete representation of the shear field when $z$ and $z'$ span $[0,\infty)$. 

\subsection{The Configuration-Space Representation}
\label{The Configuration-Space Approximation}
As an alternative to performing a cosmic shear statistic in Fourier/Bessel space the analysis can be done in
real/angular/configuration space, where instead of an angular wavenumber an angle $\theta$ is used 
on the celestial sphere as the dependent variable. Such statistics are readily computed from data by summing over
pairs of galaxies (see e.g. Kilbinger, 2015). From theory these are related
to the cosmic shear power spectra through a transform that results in two correlation functions that 
we derive in Appendix A
\ba
\label{cof0}
\xi_+(\theta,z,z') &=&\frac{1}{2\pi}\sum_{\ell}(\ell+0.5) d^{\ell}_{22}(\theta)\nn
&&[C^{SR,E}_{\ell}(z,z')+C^{SR,B}_{\ell}(z,z')]\nn
\xi_-(\theta,z,z') &=&\frac{1}{2\pi}\sum_{\ell}(\ell+0.5) d^{\ell}_{-22}(\theta)\nn
&&[C^{SR,E}_{\ell}(z,z')-C^{SR,B}_{\ell}(z,z')].
\ea
where $d^{\ell}_{22}$ and $d^{\ell}_{-22}$ are Wigner small-$d$ 
matrices\footnote{We provide tabulated values of these here {\tt http://goo.gl/UUQIUx}.}. 
$\theta$ are angular seperations on the 
sphere. This can be derived in a number of ways either starting from Hu (2000, Appendix A), from the results of Ng \& Liu 
(1999), or from considering the additive properties of the Wigner large-$D$ matrices. 
In this case the power spectra in the integrals are a combinations of both E-mode and B-mode components; however from 
theory the B-mode is typically always zero. Note that the spin nature of the field must be considered in 
relating the power spectra to the correlation functions and it should not be treated as a scalar field.

\subsubsection{Large Wavenumber Limit}
In the limit that $\ell\gg |m|$, $|m'|$ (in the cosmic shear case $|\ell|\gg 2$) the Wigner-d matrices 
can be written as Bessel functions of the first kind, 
which is what has been done in cosmic shear studies to date. Making the further 
approximation that $\ell\simeq \ell+0.5$ the transforms in equation (\ref{cof0}) are commonly 
assumed to be Hankel transforms: 
\ba
\label{cof}
\xi_+(\theta,z,z') &=&\frac{1}{2\pi}\sum_{\ell} \ell J_0(\ell\theta)\nn
&&[C^{SR,E}_{\ell}(z,z')+C^{SR,B}_{\ell}(z,z')]\nn
\xi_-(\theta,z,z') &=&\frac{1}{2\pi}\sum_{\ell} \ell J_4(\ell\theta)\nn
&&[C^{SR,E}_{\ell}(z,z')-C^{SR,B}_{\ell}(z,z')].
\ea
Hankel transforms can be performed using either a three-dimensional power spectrum, 
as we have used here, or on tomographically binned data. An inverse-Hankel transform
can also be defined e.g., $C^{SR}_{\ell}(z,z')=\int {\rm d}\theta\theta J_0(\ell\theta)\xi_+(\theta,z,z')$ but
since this formally requires an integration over \emph{all} angles it is not well-defined in a flat-sky case.

\begin{figure*}
\centering
\includegraphics[angle=0,clip=,width=2\columnwidth]{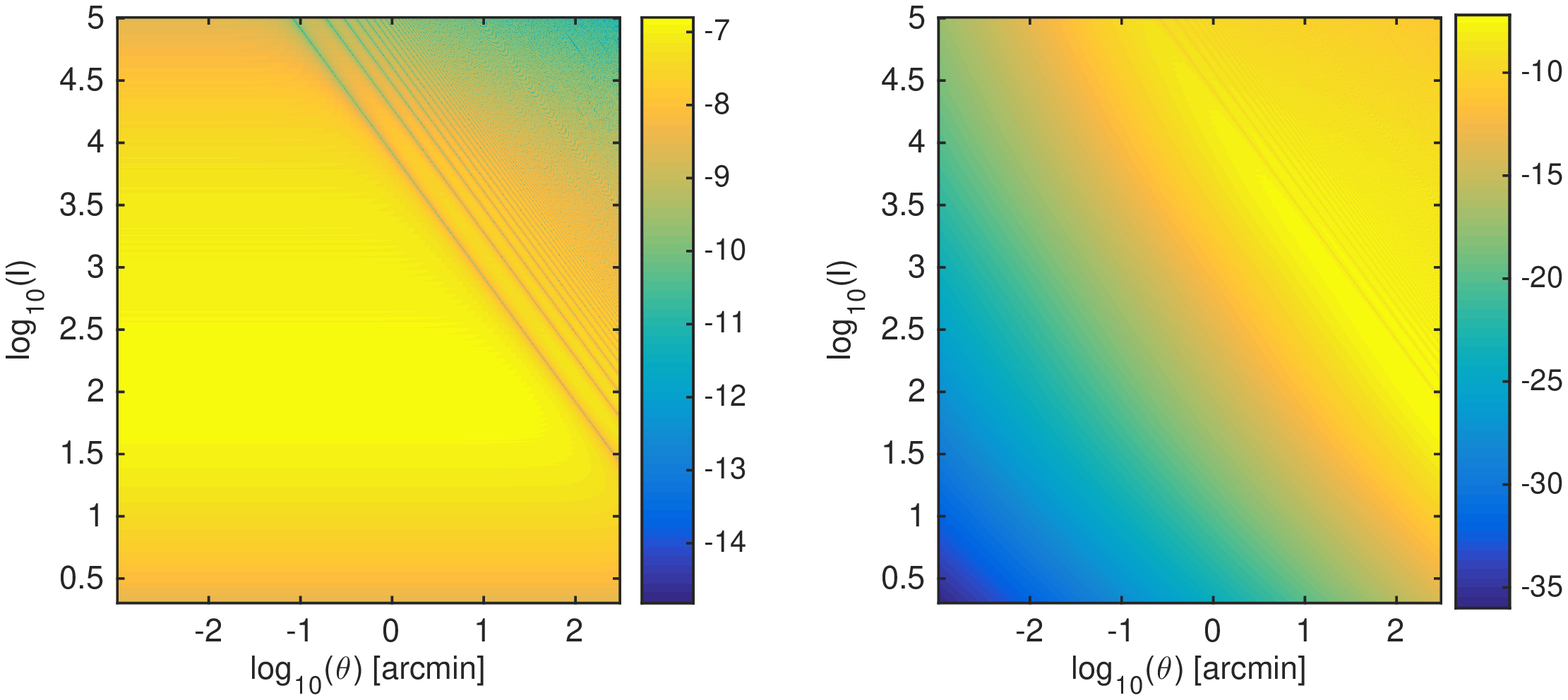}
\includegraphics[angle=0,clip=,width=2\columnwidth]{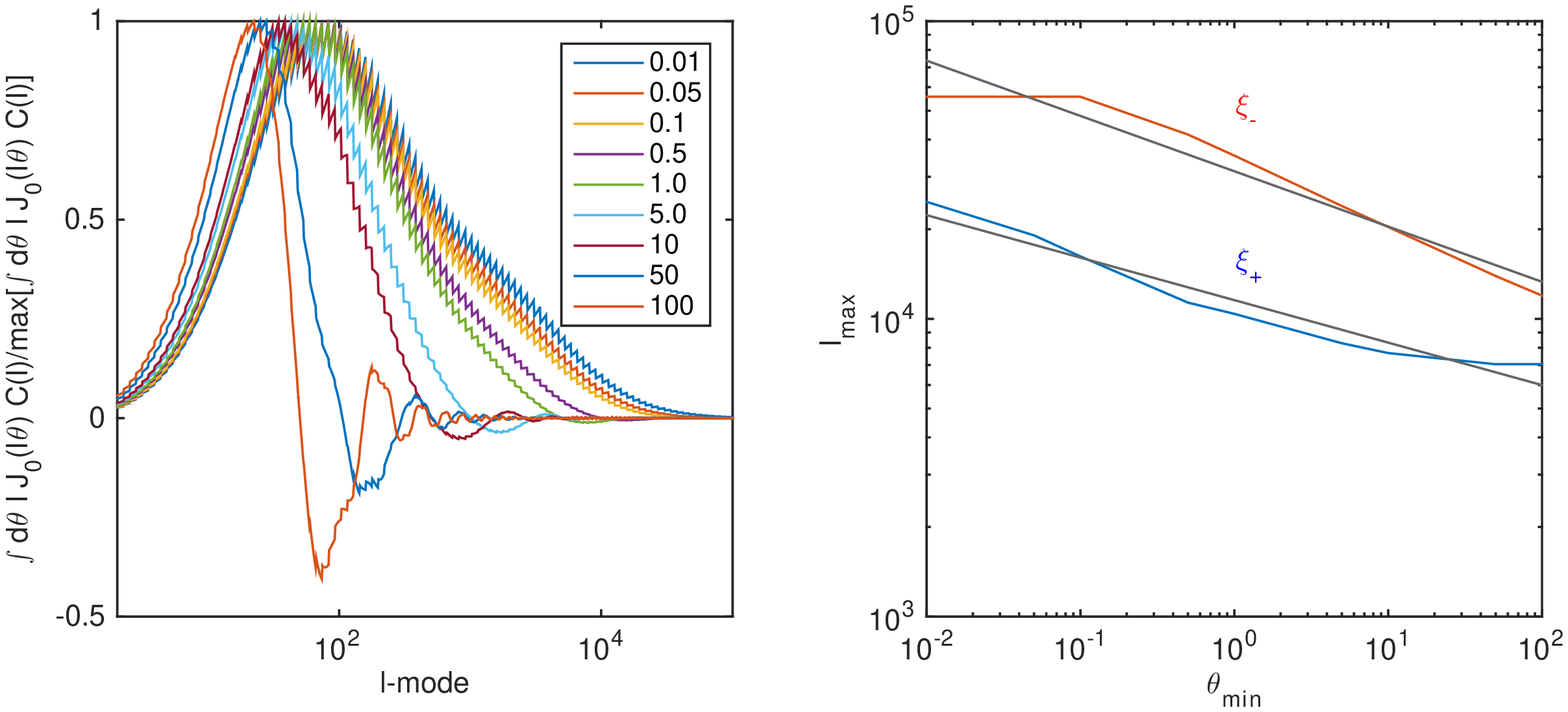}
\caption{The top panels show the functions $\ell J_0(\ell\theta)C^{SR}_{\ell}(z,z)$ (left-hand panel)
and $\ell J_4(\ell\theta)C^{SR}_{\ell}(z,z)$ (right-hand panel) for an auto-correlation
cosmic shear power spectrum $C^{SR}_{\ell}(z,z)$ evaluated at a redshift of zero; although
we find results in this Figure are insensitive to this assumption. The
colour scales denote the logarithm (base-10) of the magnitude 
of the functions that correspond to the $\xi_+$ and $\xi_-$ Hankel transforms (equation \ref{cof}) respectively. 
The lower left
panel shows the normalised integral $\int {\rm d}\theta \ell J_0(\ell\theta)C^{SR}_{\ell}(z,z)$
over $\theta$ to show the integrated weighting of $\int {\rm d}\theta \xi_{\pm}(\theta)$ as a function of $\ell$-modes
for a variety of angular ranges $\theta_{\rm min}\leq \theta/{\rm arcmin}\leq 100$. The different 
labelled colours in this plot show different
$\theta_{\rm min}$ in arcminutes. The lower right-hand panel shows the $\ell_{\rm max}$ where these integrals converge as a function of
$\theta_{\rm min}$ for the $\xi_+$ (blue) and $\xi_-$ (red) Hankel transforms. 
The fitted functions in equation (\ref{a2l}) are shown in grey.}
\label{fcof}
\end{figure*}

In the cosmic shear representations that are based on spherical harmonic transforms the angular wavenumbers can be 
approximately related to celestial
angular separations through $\theta=\pi/\ell$. However after performing the Hankel transformation the
relationship between the angle $\theta$ in equations (\ref{cof}) is more complicated. To investigate this relation we plot in
Figure \ref{fcof} the Bessel function amplitudes in equation (\ref{cof}) as a function of $\ell$-mode and $\theta$, for the
$\xi_+$ and $\xi_-$ functions.
It is clear from these figures that every angle samples from all $\ell$-modes but weighted in a different way.
To estimate which $\ell$-modes contribute to the Hankel transform integrals we compute the following integrals over $\theta$
\ba
{\mathcal W}_+(\ell,z,z')&=&\int_{\theta_{\rm min}}^{\theta_{\rm max}} {\rm d}\theta [\ell J_{0}(\ell\theta)C^{SR}_{\ell}(z,z')]\nn
{\mathcal W}_-(\ell,z,z')&=&\int_{\theta_{\rm min}}^{\theta_{\rm max}} {\rm d}\theta [\ell J_{0}(\ell\theta)C^{SR}_{\ell}(z,z')].
\ea
These are the weight functions in $\ell$-mode, integrated over all angles, that are applicable for analyses that 
require a sum over angle (such as a likelihood function).
We use $\theta_{\rm max}=100$ arcminutes, and vary $\theta_{\rm min}$ and show these functions in Figure \ref{fcof}. To compute
the maximum $\ell$-mode to minimum $\theta$ relationship we compute the cumulative functions
\be
\left(\frac{1}{A}\right)\int^{\ell_{\rm max}}_2 {\rm d}\ell\, |{\mathcal W}_+(\ell,r,r')|=f\le 1, 
\ee
that we calculate as a discrete sum, 
where $A=\int^{\infty}_2 {\rm d}\ell\,|{\mathcal W}_+(\ell,r,r')|$. These functions only
converge to machine precision at $\ell_{\rm max}\rightarrow\infty$ so in practice a tolerance needs to be defined $f$
where it is considered that most of the information is captured. We set this to $f=0.995$, i.e. $99.5\%$ of the
integral content is captured by this limit; we find that setting a limit larger than this results in 
numerical errors becoming dominant.
We plot this derived $\ell_{\rm max}$ in Figure \ref{fcof}, and find that the link between
$\ell_{\rm max}$ and $\theta_{\rm min}$ is well-approximated by the following scaling functions
\ba
\label{a2l}
\xi_+:\log_{10}[\ell_{\rm max}]&=&-0.14\log_{10}(\theta_{\rm min}/{\rm arcmin})+4.06\nn
\xi_-:\log_{10}[\ell_{\rm max}]&=&-0.19\log_{10}(\theta_{\rm min}/{\rm arcmin})+4.49.
\ea
We find that the $\xi_-$ statistic is much more sensitive to
high-$\ell$ modes than $\xi_+$. For typical minimum angles used in
data analysis of $\theta_{\rm min}\sim 0.1$ arcminutes we find that the maximum wavenumber probed is approximately
$\ell_{\rm max}\sim 5\times 10^4$ for the $\xi_+$ statistic, but the bulk of the signal comes from $\ell < 1000$.

Finally there are several ways of filtering the `raw' correlation
function measurement (equation \ref{cof}) that have been proposed for 
example Top-hat statistics, Map statistics (e.g. Munshi et al., 2004) and
COSEBIs (e.g. Schneider et al., 2010). The motivation for these, and their 
mathematical detail, are well summarised and reviewed in Kilbinger (2015).

\section{Cosmic Shear Approximations}
\label{Cosmic Shear Approximations}
We will now investigate the impact of several approximations that are commonly used in cosmic shear studies. 
We will address the flat-sky and Limber approximations, but will not discuss 
source-source clustering (Schneider et al. 2002), source-lens clustering 
(Bernardeau 1998, Hamana et al. 2002), the Born approximation (Cooray \& Hu, 2002), 
higher-order power spectrum terms (Krause \& Hirata, 2010), or unequal-time correlators (Kitching \& Heavens, 2016); 
all of which are expected to have an effect for future surveys (Euclid, LSST and WFIRST) but not for current surveys. 

\subsection{The Flat-Sky Approximation}
\label{The Flat-Sky Approximation}
The flat-sky approximation assumes that the angular extent of the observational field is small and hence the
geometry of the angular component is assumed to be planar (i.e., Euclidean). In this case a planar transform is 
done instead of a spherical transform in equations (\ref{a}) and (\ref{b}) which results in an exponential 
term ${\rm exp}({\rm i}\bell.\theta)$ instead of the spin spherical harmonics. 

In the case of computing the transform coefficients from data, equations (\ref{a}) and (\ref{b}), this results 
in a different sum over galaxies. In the computation of data vectors the weighting as a function of $\ell$ mode is therefore 
significantly different (see e.g. Hu, 2000). 

However in the computation of the theoretical covariances, due to the similar orthogonality relations between both the 
spherical harmonic and the exponential functions, equations (\ref{sb}) and (\ref{sr}), this 
only results in a simple change to the pre-factor $|D_{\ell}|$ 
from $|D_{\ell}| = \sqrt{(\ell+2)!/(\ell-2)!}\rightarrow \ell^2$. This is 
a result of the different ways that the spin raising and lowering operators (that relate the shear field to the 
gravitational potential field) act on the spin spherical harmonics and the exponential functions; 
see Appendix A of Castro et al. (2005). The impact of 
this approximation on the amplitude of the cosmic shear covariance can then very simply be computed.
It is a poor approximation as it introduces errors of order $1/\ell$, which may not be negligible. 

We note that taking a small angle approximation of the spherical
harmonics (see Castro et al., 2005 Section V; or Varshalovich, Moskalev, \& Khersonski{\u\i}, 1988 
for more complete expressions)
results in much larger differences in the amplitude of the power spectra than that 
captured in the change of local derivative of the lens potential, but this case has not been considered in the
cosmic shear literature to date.

\subsection{Tomographic Data Compression}
The tomographic approximation involves the computation of 
projected two-dimensional power spectra in a series of redshift bins including the 
inter-bin (auto-correlation) and intra-bin (cross-correlation) power spectra. This is not an approximation 
in itself, but it is a lossy data compression.

We look at the effect of this binning by first relating the spherical-Bessel and spherical-radial transforms together. 
As shown in Kitching et al. (2014) the shear transform coefficients, from our equations (\ref{a}) and (\ref{b}), can be 
related through a radial transform 
\be
\label{ss}
\gamma^m_{\ell}(z_1)=\int {\rm d}r W^{SR}[z_1,z(r)]\int {\rm d}k j_{\ell}(kr)\gamma^m_{\ell}(k)
\ee
where the weight function is the same one that appears in equation (\ref{Gsr}), where the integrand of comoving 
distance $r$ is related to a redshift $z(r)$, and describes the bins as a function of redshift. When 
referring to tomography we use numbered redshifts e.g. $z_1$ and $z_2$, rather than $z$ and $z'$. We note that 
only in the case that the weight function is a delta-function is this a full description of the three-dimensional shear field. 
In the case that the bin-width is finite we will refer to this as a `tomographic' representation of the shear field.

By taking the covariance of equation (\ref{ss}) the two power spectra can be related through 
\ba
\label{sb2sr}
C^{SR}_{\ell}(z_1,z_2)=&&\int {\rm d}k{\rm d}k'{\rm d}r'{\rm d}r''\nn
&&W^{SR}[z_1,z(r')]W^{SR}[z_2,z(r'')]\nn
&&j_{\ell}(kr')j_{\ell}(k'r'')C^{SB}_{\ell}(k,k'). 
\ea
This transformation from spherical-Bessel to spherical-radial (tomographic) representations 
can be performed for any integrable weight function $W^{SR}$; this is also discussed in Castro et al. (2005).  

The reverse transform can also be computed, but \emph{only} in the case that the weight function 
is a delta-function in redshift. In 
this specific case the reverse transform is 
\be 
\label{sr2sb}
C^{SB}_{\ell}(k,k')=\int{\rm d}z{\rm d}z' j_{\ell}(kr[z])j_{\ell}(k'r[z'])C^{SR}_{\ell}(z_1,z_2), 
\ee
where the integration over redshift is formally over $0\leq z < \infty$.

It has been shown (e.g. Bridle \& King, 2007) that, because of intrinsic alignments, $10-20$ redshift bins are required 
in order for the cosmic shear power spectrum 
to be sufficiently sampled in redshift to extract the majority of cosmological information. 
This is because the lensing kernel is 
a relatively broad function in redshift space. This is applicable when describing the shear field using the spherical-radial 
representation, with the caveats that such current studies of the 
convergence of this approximation have assumed the flat-sky and Limber approximations (that we discuss in the next Section). 

\subsection{The Limber Approximation}
\label{The Limber Approximation}
The Limber (Limber, 1953) approximation was first introduced in Kaiser (1998) for cosmic shear studies 
as a method for rendering the calculations more tractable and understandable, and has subsequently been used in the 
majority of the cosmic shear studies, both in methodological development and in applications to data. 
In LoVerde \& Afshordi (2008) a particularly clear explanation of the approximation was provided. This assumed 
that the matter power spectrum was not evolving, i.e. it can be expressed as a function of $k$-mode only 
$P(k)$ (LoVerde \& Afshordi, 2008; equation 5). Unfortunately the LoVerde \& Afshordi (2008) approximation 
is not directly appropriate at all orders for the cosmic shear setting where the shear field is 
an integrated effect over an evolving matter power spectrum; an assumption that we address in Appendix B.   
In Kitching, Heavens, Miller (2011) the effect of the Limber approximation on cosmic shear was investigated using 
the LoVerde \& Afshordi (2008) approximation, and an effect on the expected error bars of 
cosmological parameters was predicted. 

If the Limber approximation is assumed then using the Kaiser (1998) and LoVerde \& Afshordi (2008) approximation 
the spherical-radial representation of the cosmic shear field can be written as 
\ba
\label{lim}
C^{SR}_{\ell}(z_1,z_2)\simeq |D_{\ell}|^2{\mathcal A}^2 
\int \frac{{\rm d}k}{k^2} P(k,\nu/k)f(z_1,\nu, k)f(z_2,\nu, k)
\ea
where the variable $\nu=\ell+1/2$. In Appendix B we show that this is indeed the first order approximation to the 
cosmic shear power spectrum despite the assumption of a constant matter power spectrum, 
however the expansion of this to higher order results
in a convergence towards the unapproximated case 
only if redshift-independent limits in angular wavenumber are assumed. The kernel functions are 
\ba
f(z_1,\nu, k)=&&\left(\frac{1}{\nu k^2}\right)^{1/2}\int {\rm d}z'{\rm d}z_p n(z_p)p(z'|z_p)W^{SR}(z_1,z_p) \nn
&&\frac{F_K(r[z'],\nu/k)}{a(\nu/k)}.
\ea
This expression is not entirely in the same form as commonly used in the cosmic shear literature (e.g. Hu, 1999; 
Joachimi \& Bridle, 2010; Heymans et al. 2013), where the standard form is to use an inner integral 
over $r$ instead of $k$-mode. 
As shown in Appendix B when doing this we find that the Limber-approximated power is given by 
\be 
C^{SR,L}_{\ell}(z_1, z_2)\simeq |D_{\ell}|^2{\mathcal A}^2 \left(\frac{1}{\nu^4}\right)
\int {\rm d}r \frac{q(r_1,r)q(r_2,r)}{r^2} P(\nu/r,r).
\ee
where
\ba
q(r_1,r)=&&\frac{r}{a(r)}\int {\rm d}z_p{\rm d}z'n(z_p)p(z'|z_p)W^{SR}(z_1,z_p)\nn
&&\left(\frac{r(z')-r}{r(z')}\right)
\ea
where we have expanded the function $F_K$, and we have assumed here a flat-geometry ($K=0$). 
This is the standard form for the cosmic shear power spectrum (see e.g. Hu, 1999; Joachimi \& Bridle, 2010), except that
there is an $\ell$-dependent pre-factor 
\ba
\label{tl}
T_{\ell}=\frac{|D_{\ell}|^2}{\nu^4}=\frac{(\ell+2)(\ell+1)\ell(\ell-1)}{(\ell+0.5)^4}.
\ea
$T_\ell$ is normally replaced by 1.  One justification for this is to replace the numerator by $\ell^4$ in the flat-sky approximation, and to take a high-$\ell$ approximation $\nu\simeq\ell$ in the denominator.   Note that a flat-sky approximation that also retains the Limber $\nu^{-4}$ dependence would lead to an inaccurate $T_\ell$ which differs from unity at ${\mathcal{O}}(1/\ell)$, and leads to significant errors at low $\ell$.  Note that $T_\ell$ differs from unity only at ${\mathcal{O}}(1/\ell^2)$, so the standard approximation is good for current data, but there is no reason at all not to use the full expression.  

Up to first order the Limber approximation can be summarised by comparing equation (\ref{sr}) with equation (\ref{lim}) as a 
replacement of Bessel functions with scaled delta functions inside the integrals
\be 
\label{bl}
j_{\ell}(kr)\rightarrow \sqrt{\frac{\pi}{2\ell+1}}\delta^D(\ell+1/2-kr).
\ee
This expression shows how the Limber approximation acts to link the angular and 
radial modes through the relation $\ell=kr[z]-1/2$, that we also 
derive in Appendix B, which has an important effect on the computation of cosmic shear power spectra.

\subsection{The Impact of the Approximations} 
\label{The Impact of the Approximations}
There are various steps in the derivation of a configuration-space shear statistic, which involve relating the lensing potential power spectrum on the (spherical) sky to the matter power spectrum, then computing the shear power spectrum on the sky, and from there transforming to configuration space if desired.   These steps can introduce approximations beyond the Born approximation and approximations of unequal time correlators, but some are not necessary.   At the first stage, it may be necessary to use the Limber approximation for computational tractability reasons. At low $\ell$ this is a poor approximation, and if speed is an issue, the next term in the Limber approximation (LoVerde \& Afshordi 2008) should be considered.  In moving from lensing potential to shear, the full $\ell$-dependent prefactor of $(\ell+2)(\ell+1)\ell(\ell-1)$ should be included, and not approximated by the flat-sky $\ell^4$ value.  If the Limber approximation is used, $\ell+1/2$ should not approximated by $\ell$. Finally, in computing configuration-space quantities such as shear
correlation functions, finite sums over $\ell$ should be done, using Wigner small-d matrices (equation \ref{cof0}), and not approximated by Hankel transforms.

In summary in going from the full cosmic shear expressions to those that are commonly used there are a series of 
approximations. These are, starting from a spherical-sky non-Limber-approximated power spectrum: 
\begin{itemize}
\item 
{\bf Flat-Sky Approximation:} The assumption of a flat-sky changes the pre-factor in the shear-shear power spectrum from 
$(\ell+2)(\ell+1)\ell(\ell-1)$ to $\ell^4$. This is inaccurate and unnecessary.
\item 
{\bf Limber Approximation:} The first-order Limber approximation involves changing the Bessel functions to scaled delta functions using equation (\ref{bl}), leading to a prefactor in the shear power spectrum of
$(\ell+2)(\ell+1)\ell(\ell-1)/(\ell+0.5)^4$. 
\item 
{\bf Prefactor Unity Approximation:} In the Limber function expression a further approximation can be made that the 
$\ell$-dependent pre-factor is unity i.e. $T_{\ell}=1$ in equation (\ref{tl}).  This is good to ${\mathcal O}(1/\ell^2)$, but is unnecessary.
\item 
{\bf Integral Variable Approximation:} In the Limber approximation the inner variable $\ell+0.5$ is sometimes replaced by $\ell$ in the argument to the matter power spectrum. This is inaccurate and unnecessary and is not used in this paper.
\item 
{\bf Hankel Transform Approximation:} Then when transforming to real-space a Hankel transform can be used instead of a 
spherical sky correlation function (equation \ref{cof0}).  This leads to an increasing error with angle, and a spherical summation over $\ell$ modes is preferred.
\end{itemize} 
Each of these approximations act independently, the first four act on the cosmic shear power spectrum, and the last only 
in the case that this is transformed to real-space.

\subsubsection{Impact on the Power Spectrum}
\begin{figure*}
\centering
\includegraphics[angle=0,clip=,width=2\columnwidth]{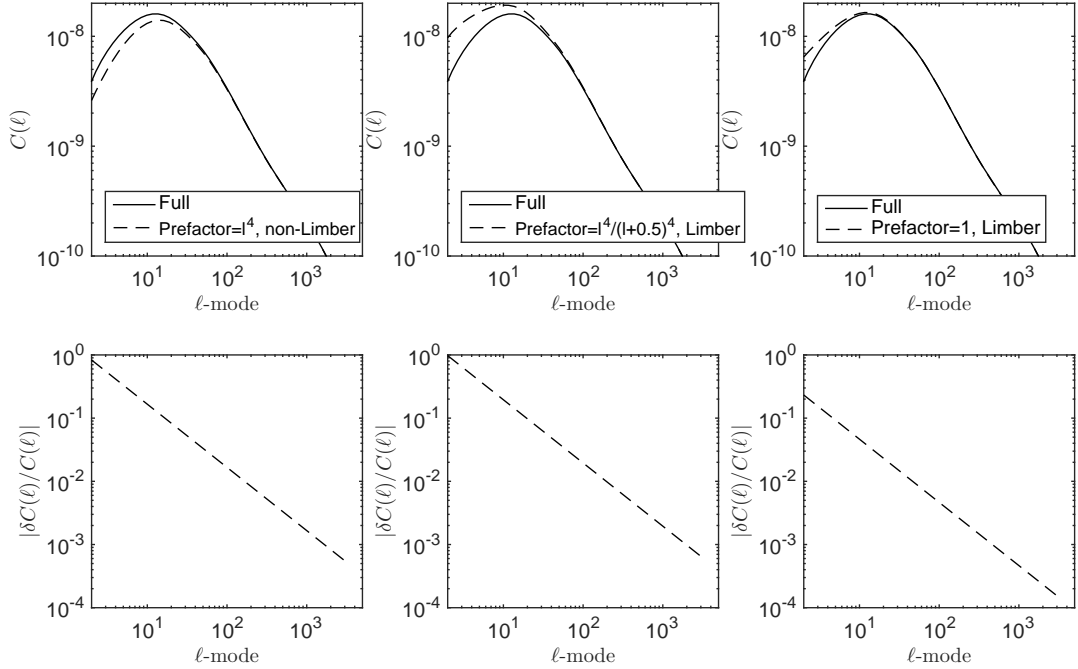}
\caption{Top panels: The solid line is the full $C(\ell)$ cosmic shear power spectrum, for a CFHTLenS
$n(z)$; not assuming any of the approximations listed 
in Section \ref{The Impact of the Approximations} i.e. flat-sky, Limber, prefactor-unity or integral variable assumptions. 
In the full case the $\ell$-dependent prefactor to the power spectrum is $(\ell+2)!/(\ell-2)!$ and the Limber 
approximation is not assumed. 
The dashed lines show the power spectrum when each of the approximations is applied in combination in the panels from left to right, the $\ell$ expressions denote the power spectrum pre-factor used. 
The lower panels show the modulus of the fractional difference between the 
full case and the approximated cases $|[C^{\rm Full}(\ell)-C^{\rm Approx}(\ell)]/C^{\rm Full}(\ell)|$.}\label{tl_effect}
\end{figure*}

In Figure \ref{tl_effect} we show the impact of the Flat-Sky, Limber and Prefactor-Unity approximations. Throughout we 
do not make the Integral Variable Approximation, and use a cosmology equal to the Planck, (2016; Table 4 TT+low P) best fit values. 
It can be seen that for $\ell\ls 10$ there is a more than $10\%$ suppression in the power due to the 
Flat-Sky Approximation which reduces to $\ls 1\%$ for $\ell\gs 100$.
We can assess the impact of these approximations by computing the integrated effect over the differences
\be
\label{barA}
\langle {\mathcal A}\rangle/N_A=
\frac{\int {\rm d}\ln\ell\,\ell^2 \delta C(\ell)}{\int {\rm d}\ln\ell\,\ell^2}
\ee
complementary formulations are provide for this quantity in Massey et al. (2013), Cropper et al. (2013)
and Amara \& Refregier (2008); here we include a normalisation 
$N_A=\int {\rm d}\ln\ell\,\ell^2$ as suggested by Massey et al. (2013). 
In general a non-zero $\langle {\mathcal A}\rangle$
will change the amplitude of the power spectrum and bias cosmological parameter inference.
As discussed in Massey et al. (2013) the requirement on the amplitude of this
quantity is $\langle {\mathcal A}\rangle/N_A\leq 1.8\times 10^{-12}$ for a Euclid- or LSST-like weak lensing 
survey to return unbiased results on the dark energy equation of state parameters,
this requirement is an allowance for \emph{all} systematic effects including instrumental and algorithmic quantities.
We find that for best approximated case to the full power spectrum (where the prefactor is unity and the Limber 
approximation is assumed) that $\langle {\mathcal A}\rangle/N_A=1.9\times 10^{-13}$, that would account for 
$11\%$ of the total budget for systematic effects for a Euclid or LSST-like experiment that suggests 
such approximations should not be used. If scales of $\ell<100$ are ignored then we find only a 
modest change with $\langle {\mathcal A}\rangle/N_A=1.7\times 10^{-13}$ (note the $\ell^2$ factor in 
equation \ref{barA} that gives higher weight to larger $\ell$-modes). 

\subsubsection{Impact on the Correlation Functions} 
\begin{figure*}
\centering
\includegraphics[angle=0,clip=,width=2\columnwidth]{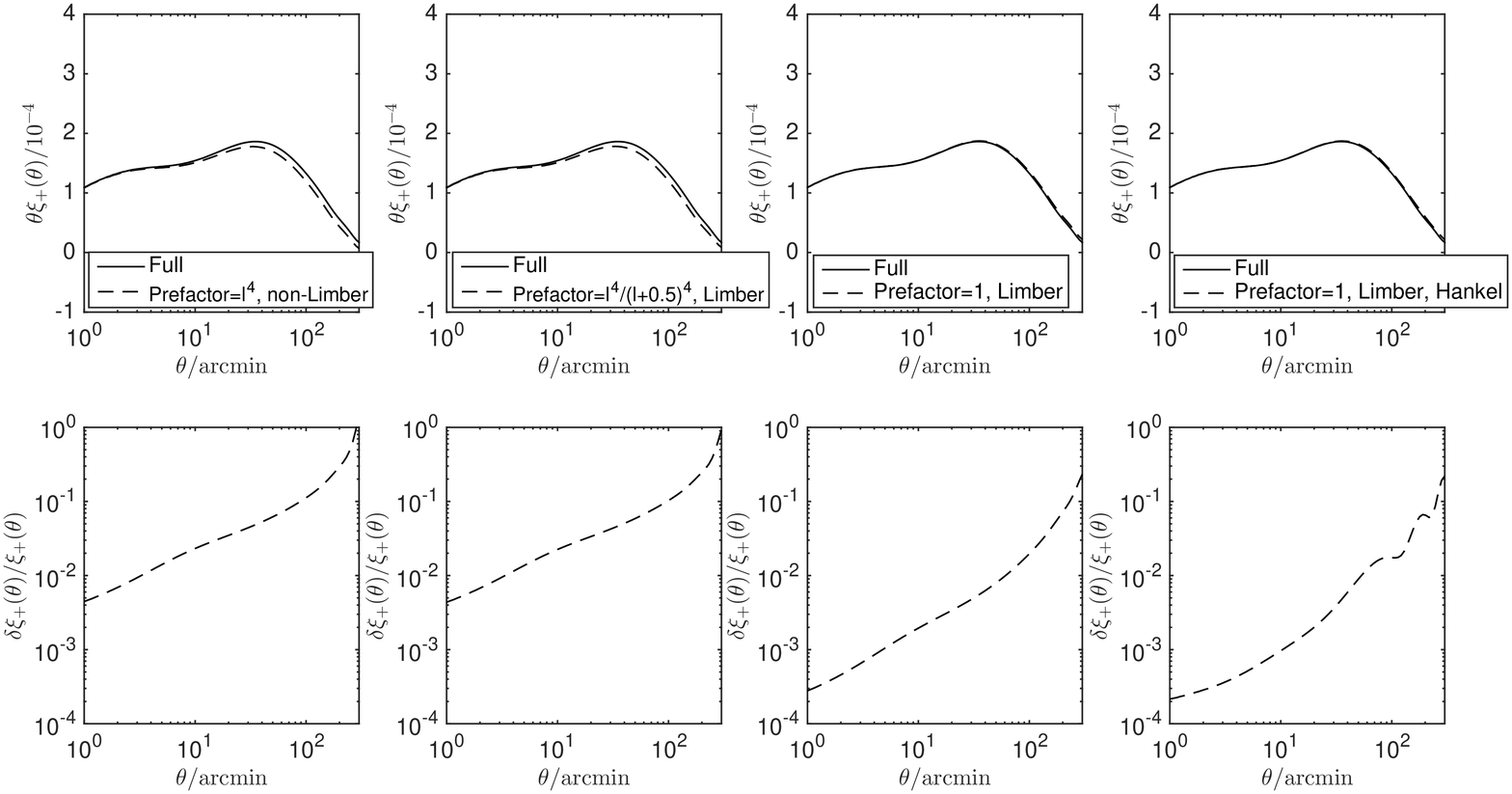}
\caption{Top panels: The solid line is the full projected $\xi_+(\theta)$ cosmic shear correlation function, for a CFHTLenS 
$n(z)$; 
not assuming any of the approximations listed
in Section \ref{The Impact of the Approximations} i.e. flat-sky, Limber, prefactor-unity, integral variable, or 
Hankel assumptions. In the full case the $\ell$-dependent prefactor to the power spectrum is 
$(\ell+2)!/(\ell-2)!$, the Limber approximation is not assumed, and a transform using Wigner small-d matrices 
(equation \ref{cof0}) is used. 
The dashed lines show the 
correlation function when each of the approximations is applied in combination in the panels from left to right. The 
lower panels show the modulus of the fractional different between the full case and the approximated cases 
$|[\xi_+^{\rm Full}(\theta)-\xi_+^{\rm Approx}(\theta)]/\xi_+^{\rm Full}(\theta)|$.} 
\label{xp_effect}
\end{figure*}

\begin{figure*}
\centering
\includegraphics[angle=0,clip=,width=2\columnwidth]{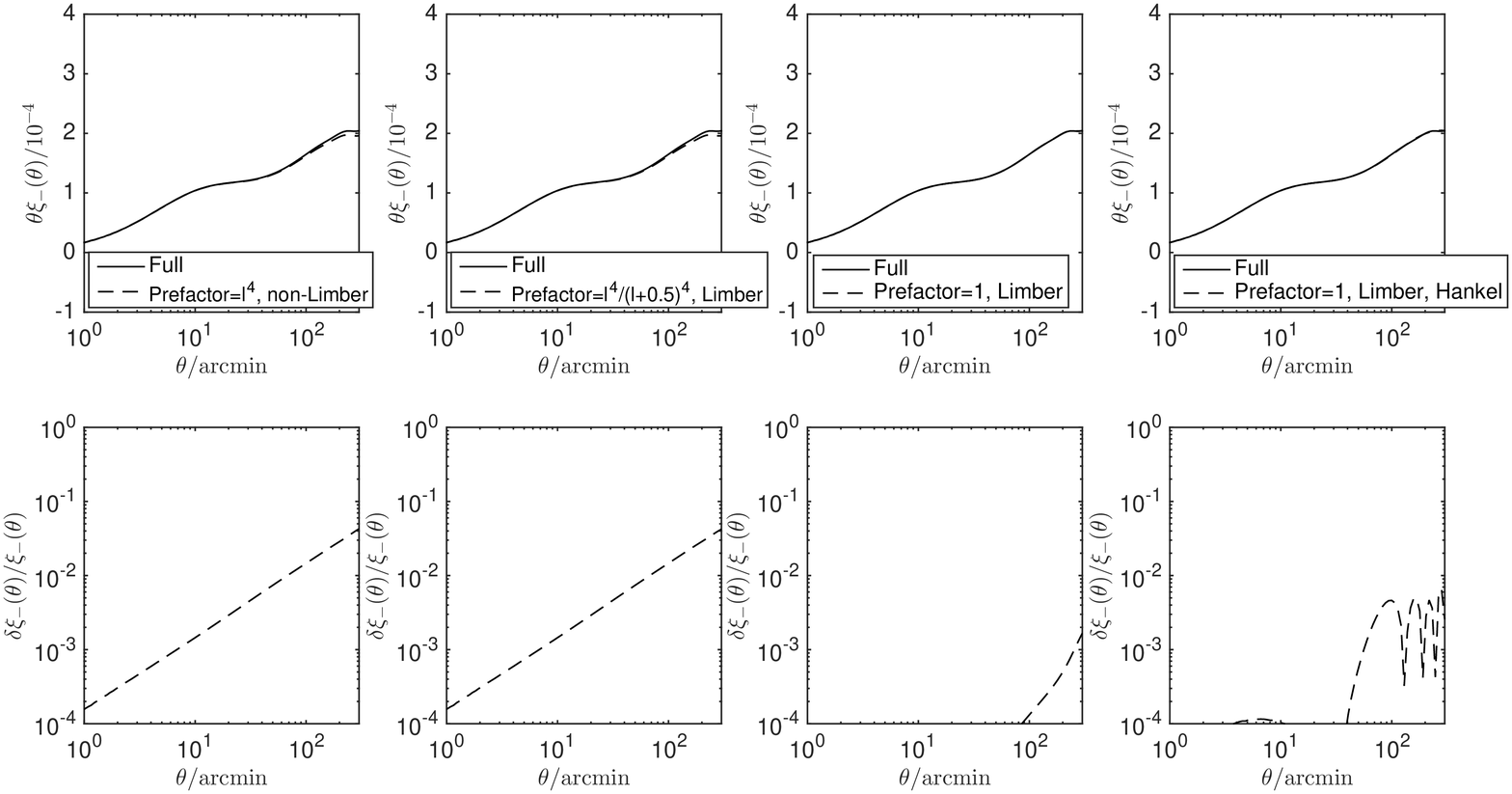}
\caption{Top panels: The solid line is the full projected $\xi_-(\theta)$ cosmic shear correlation function, for a CFHTLenS
$n(z)$;
not assuming any of the approximations listed
in Section \ref{The Impact of the Approximations} i.e. flat-sky, Limber, prefactor-unity, 
integral variable, or Hankel assumptions. 
In the full case the $\ell$-dependent prefactor to the power spectrum is
$(\ell+2)!/(\ell-2)!$, the Limber approximation is not assumed, and a transform using Wigner small-d matrices
(equation \ref{cof0}) is used.
The dashed lines show the
correlation function when each of the approximations is applied in combination in the panels from left to right. The
lower panels show the modulus of the fractional different between the full case and the approximated cases
$|[\xi_-^{\rm Full}(\theta)-\xi_-^{\rm Approx}(\theta)]/\xi_-^{\rm Full}(\theta)|$.}
\label{xm_effect}
\end{figure*}

In Figure \ref{xp_effect} and \ref{xm_effect} we show the impact of the successive approximations on the real-space correlation 
functions. 
\begin{figure*}
\centering
\includegraphics[angle=0,clip=,width=1.5\columnwidth]{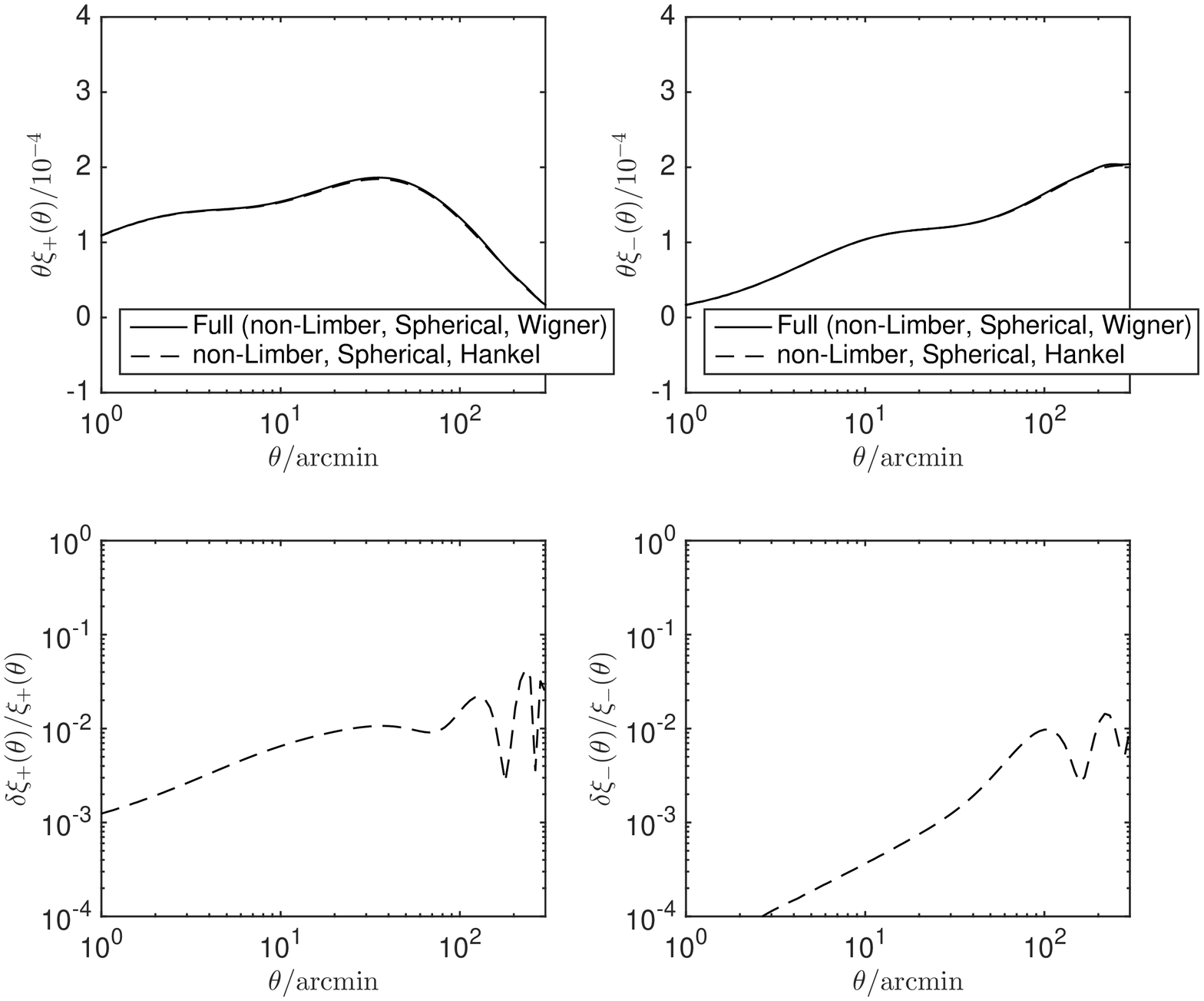}
\caption{Top panels: The solid line is the full projected $\xi_{+/-}(\theta)$ cosmic shear correlation function, 
for a CFHTLenS $n(z)$;
not assuming any of the approximations listed
in Section \ref{The Impact of the Approximations} i.e. flat-sky, Limber, prefactor-unity,
integral variable, or Hankel assumptions. The dashed lines show the
power spectrum when the Hankel transform instead of the full Wigner-d expression is used. The 
lower panels show the modulus of the fractional different between the full case and the approximated cases.}
\label{wh_effect}
\end{figure*}

Similarly to the power spectrum investigations we find that the Flat-sky approximation on its own has a large effect, but that 
again the assumption of a unity prefactor cancels out the 
approximation changes somewhat. In general we find that these low-$\ell$ approximations have a more significant impact 
on $\xi_+$ than $\xi_-$, as may be expected from Section \ref{The Configuration-Space Approximation}. 
The additional step of assuming a Hankel transform rather than a transform 
that uses Wigner small-d matrices (equation \ref{cof} instead of equation \ref{cof0}) 
results in only a small additional change at scales greater than $10$ arcminutes; we show only this effect in 
Figure \ref{wh_effect}. 

There are currently no explicit requirements 
set on the correlation function amplitude changes in the literature for future experiments that we are aware of, so it 
is not possible to assess the applicability of these requirements for \emph{Euclid}-like experiments. However we note that 
percent to tens of percent-level changes can occur and, given that the full case is not particularly more computationally 
demanding than the approximate cases, we recommend that the full case is used. 

\subsection{A Schema of Cosmic Shear Statistics}
Each of the cosmic shear representations and approximations can be linked in a series of transformations that relate one 
to the other. For example in Kitching, Heavens, Miller (2011) and Kitching et al. (2014) we show how 
to relate the spherical-Bessel to the tomographic representation (we also show this in Appendix B). In 
this paper we show how to transform from the spherical-Bessel to spherical-radial cases. The flat-sky 
and configuration-space approximations are well-known as we have discussed. 

We show how all of these are linked together in Figure 4 where we relate each of the cosmic shear 
statistics together via the network of approximations that 
can be employed. In this Figure arrows indicate the direction that the 
transform takes the statistic, where only one such case is reversible\footnote{By reversible we 
mean that it can be performed in either direction, without loss of information.} (the three-dimensional radial transform). 
We also link the points at which estimators from data are linked to the theoretical statistics, 
and highlight those statistics that have been applied to data. This provides a visual way to understand what transformation need to be made to interpret any given cosmic shear data analysis, where any statistical assumptions have been made, and how a given observation can be translated into another. 
\tikzstyle{block} = [rectangle, draw, fill=blue!20,
text width=9em, text centered, rounded corners, minimum height=4em]
\tikzstyle{block2} = [rectangle, draw, fill=yellow!20,
text width=15em, text centered, rounded corners, minimum height=4em]
\tikzstyle{block2a} = [rectangle, draw, fill=yellow!20,
text width=5em, text centered, rounded corners, minimum height=4em]
\tikzstyle{block2b} = [rectangle, draw, fill=yellow!20,
text width=5em, text centered, rounded corners, minimum height=2em]
\tikzstyle{blocka} = [rectangle, draw, fill=blue!20,
text width=5em, text centered, rounded corners, minimum height=2em]
\tikzstyle{block3a} = [rectangle, draw, fill=green!20,
text width=5em, text centered, rounded corners, minimum height=2em]
\tikzstyle{block3} = [rectangle, draw, fill=green!20,
text width=9em, text centered, rounded corners, minimum height=4em]
\tikzstyle{line} = [draw,  -latex',double]
\tikzstyle{dline} = [draw, latex'-latex',double]
\tikzstyle{datal} = [draw, -latex',double]
\tikzstyle{cloud} = [draw, ellipse,fill=red!20, node distance=3cm,
minimum height=4em]
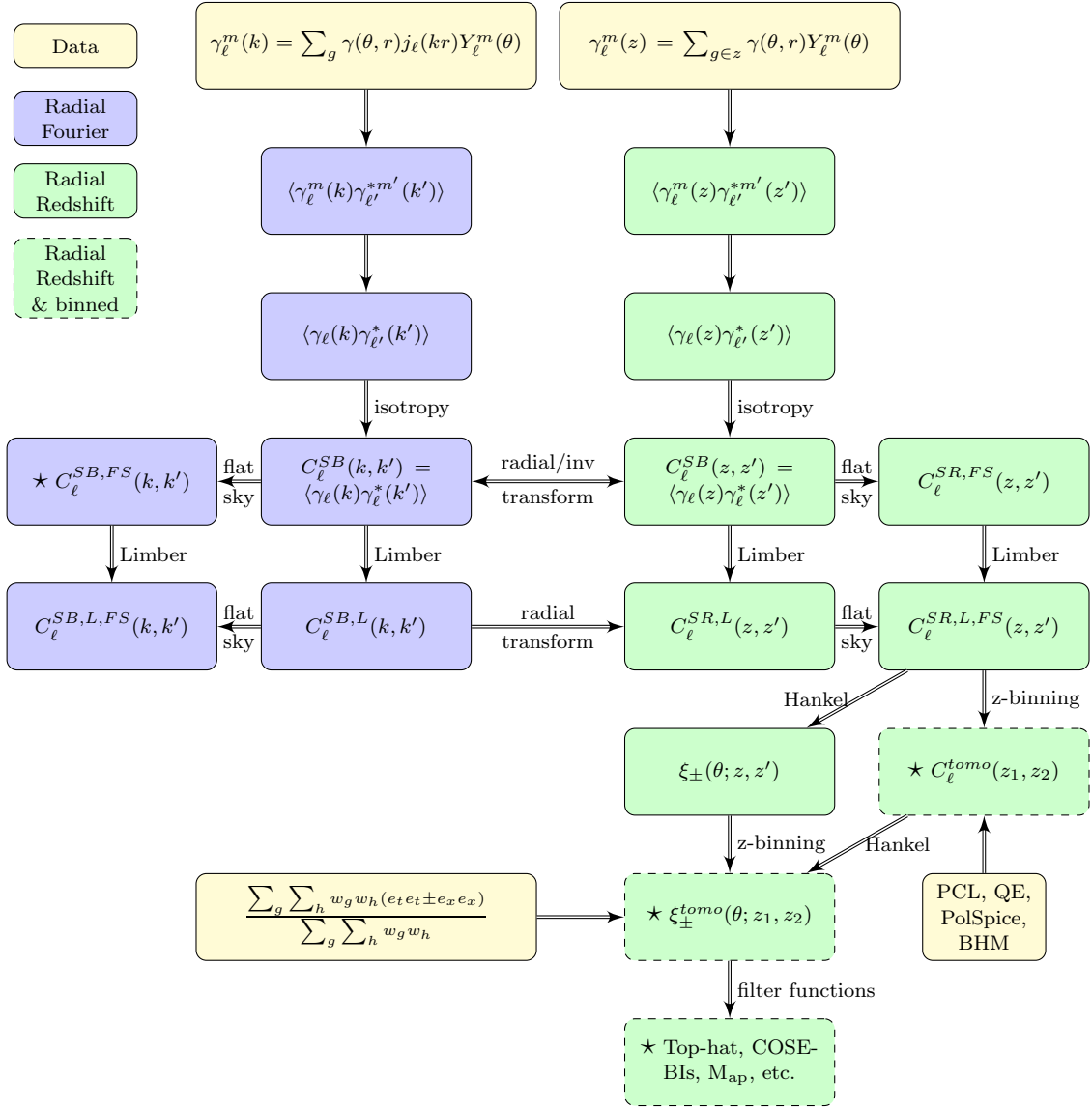
\begin{figure*}
\label{flow}
  \makebox[\columnwidth]{
  \begin{tikzpicture}[node distance = 2cm, auto,remember picture]
    \node [block2b] (l1) {Data};
    \node [blocka, below of = l1, node distance = 1cm] (l2) {Radial Fourier};
    \node [block3a, below of = l2, node distance = 1cm] (l3) {Radial Redshift};
    \node [block3a, below of = l3, node distance = 1.2cm,dashed] (l4) {Radial Redshift \& binned};

    \node [block2, right of =l1, node distance = 4cm] (init) {$\gamma^m_{\ell}(k)=\sum_g\gamma(\theta,r)j_{\ell}(kr)Y^m_{\ell}(\theta)$};
    \node [block, below of = init] (power) {$\langle\gamma^m_{\ell}(k)\gamma^{*m'}_{\ell'}(k')\rangle$};
    \node [block, below of = power] (homo) {$\langle\gamma_{\ell}(k)\gamma^{*}_{\ell'}(k')\rangle$};
    \node [block, below of = homo] (iso) {$C^{SB}_{\ell}(k,k')=\langle\gamma_{\ell}(k)\gamma^{*}_{\ell}(k')\rangle$};
    \node [block, left of = iso, node distance = 3.5cm] (fs) {{\Large $\star$} $C^{SB,FS}_{\ell}(k,k')$};
    \node [block, below of = iso] (lim) {$C^{SB,L}_{\ell}(k,k')$};
    \node [block, left of = lim, node distance = 3.5cm] (fsl) {$C^{SB,L,FS}_{\ell}(k,k')$};
    \node [block3, right of = lim, node distance = 5cm] (srlim) {$C^{SR,L}_{\ell}(z,z')$};
    \node [block3, above of = srlim] (isosr) {$C^{SB}_{\ell}(z,z')=\langle\gamma_{\ell}(z)\gamma^{*}_{\ell}(z')\rangle$};
    \node [block3, above of = isosr] (homosr) {$\langle\gamma_{\ell}(z)\gamma^{*}_{\ell'}(z')\rangle$};
    \node [block3, above of = homosr] (powersr) {$\langle\gamma^m_{\ell}(z)\gamma^{*m'}_{\ell'}(z')\rangle$};
    \node [block2, above of = powersr] (init2) {$\gamma^m_{\ell}(z)=\sum_{g\in z}\gamma(\theta,r)Y^m_{\ell}(\theta)$};
    \node [block3, right of = isosr, node distance = 3.5cm] (fssr) {$C^{SR,FS}_{\ell}(z,z')$};
    \node [block3, right of = srlim, node distance = 3.5cm] (fssrl) {$C^{SR,L,FS}_{\ell}(z,z')$};
    \node [block3, below of = fssrl,dashed] (fssrlzfs) {{\Large $\star$} $C^{tomo}_{\ell}(z_1,z_2)$};
    \node [block3, below of = srlim] (corrf) {$\xi_{\pm}(\theta; z,z')$};
    \node [block3, below of = corrf,dashed] (tcorrf) {{\Large $\star$} $\xi^{tomo}_{\pm}(\theta; z_1,z_2)$};
    \node [block3, below of = tcorrf,dashed] (cd) {{\Large $\star$} Top-hat, COSEBIs, ${\rm M}_{\rm ap}$, etc.};
    \node [block2a, below of = fssrlzfs] (tm) {PCL, QE, PolSpice, BHM};
    \node [block2, left of = tcorrf, node distance = 5cm] (cf) {$\frac{\sum_g\sum_h w_g w_h (e_t e_t \pm e_x e_x)}{\sum_g\sum_h w_g w_h}$};
    \path [datal] (init) -- (power);
    \path [line] (power) -- (homo);
    \path [line] (homo) -- node {isotropy}(iso);
    \path [line] (iso) -- node {Limber}(lim);
    \path [line] (iso) -- node[above] {flat} node[below] {sky}(fs);
    \path [line] (lim) -- node[above] {flat} node[below] {sky}(fsl);
    \path [line] (lim) -- node[above] {radial} node[below] {transform}(srlim);
    \path [dline] (iso) -- node[above] {radial/inv} node[below] {transform}(isosr);
    \path [datal] (init2) -- (powersr);
    \path [line] (powersr) -- (homosr);
    \path [line] (homosr) -- node {isotropy}(isosr);
    \path [line] (isosr) -- node[above] {flat} node[below] {sky}(fssr);
    \path [line] (srlim) -- node[above] {flat} node[below] {sky}(fssrl);
    \path [line] (fs) -- node {Limber}(fsl);
    \path [line] (fssr) -- node {Limber}(fssrl);
    \path [line] (isosr) -- node {Limber}(srlim);
    \path [line] (fssrl) -- node[left] {Hankel}(corrf);
    \path [line] (fssrlzfs) -- node[right] {Hankel}(tcorrf);
    \path [line] (corrf) -- node {z-binning}(tcorrf);
    \path [line] (tcorrf) -- node {filter functions}(cd);
    \path [line] (fssrl) -- node {z-binning}(fssrlzfs);
    \path [datal] (tm) -- (fssrlzfs);
    \path [datal] (cf) -- (tcorrf);
  \end{tikzpicture}
  }
\caption{A schema relating each cosmic shear approximation to all others. For brevity we do not include spherical-sky, 
or non-Limber-approximated Hankel-like transforms (defined in Castro et al., 2005, Section V) 
as these are not currently in use. Each arrow shows the direction in which a functional approximation or 
change is applied where the majority of approximations are irreversible. The yellow boxes show places where a statistic can, and has, 
been approximated from data; these are a direct spherical-Bessel transform (top), 
a direct correlation function estimator (left), and several 
power spectrum estimators that are Pseudo-$C(\ell)$ (e.g. Hikage et al., 2011), Quadratic Estimator (Hu \& White, 2001), 
Polspice (The DES Collaboration et al., 2015) and Bayesian Hierarchical Modelling (BHM, e.g. Alsing et al., 2016, 2017). 
The blue boxes show statistics that treat the radial (redshift) direction using a Fourier-like/spherical-Bessel analysis, the 
green boxes show statistics that treat the radial direction directly in redshift space. The 
solid framed boxes denote full three-dimensional 
statistics and the dashed framed boxes show redshift binned or `tomographic' statistics. The superscript 
acronyms SB, SR, FS and L refer to spherical-Bessel, spherical-radial, flat-sky and Limber approximations respectively. Isotropy 
refers to angular isotropy but not radial, as the shear field probes different look-back times in the expansion history.
The labels $(k$, $k')$ and $(z$, $z')$ denote continuous scale and redshift variables, and $(z_1,z_2)$ to discretised redshifts. 
The lower box denoting filter functions refers to 
Map (e.g. Munshi et al., 2004), COSEBI (Schneider et al. 2010) and top-hat statistics. The stars ${\Large \star}$ show which statistics have been applied to data.}
\end{figure*}

\section{Discussion}
\label{Discussion}
There have been several other investigations into the Limber approximation. For example 
Giannantonio et al., (2012)
concluded that the Limber approximation is accurate for $\ell\gs 20$. However  
Giannantonio et al. (2012), equations 25 and 26, 
neglect a factor of $(\ell+2)!/(\ell-2)!$ (or $l^4$ in the flat sky limit), 
and also use $k^2$ in the inner integral ($\beta$ in their notation) not $(1/k^2)$ (which is 
the appropriate factor for the cosmic shear case). 

Jeong et al. (2009) tested the effect of the Limber approximation on
the convergence-convergence power spectra and found a $\sim 1\%$ change in power at
$\ell\ls 100$, and a $10\%$ change at scales $\ell\ls 10$.
This result is partly consistent with our analysis where a $10\%$ change in the amplitude of the
$C^{SR}_{\ell}(z,z')$ shear-shear power spectrum at $\ell\sim 10$ would 
propagate into $\xi_+$ and $\xi_-$ statistics with a similar
decrease in power on the real-space angular scales presented in current data analyses. However the
range of $k$-modes and redshift ranges is not quoted in that paper (in particular if a $kr<\ell$ limit is imposed or not)
which makes a detailed comparison difficult.
Simon (2007) performed a similar study of the Limber approximation in the galaxy clustering context and 
found that there is a $\sim 10\%$ bias in the correlation function at scales of $\theta\simeq 260$ arcminutes. 
Bernardeau et al. (2012) show that the  Limber-approximated 
power spectrum is accurate to better than $1\%$ at $\ell > 8$, however their non-Limber approximated expression uses 
the primordial Newtonian potential power spectrum $P(k)$ that is non-evolving (see their equation 45 where the power 
spectrum is taken out of the integrations over comoving distance). 

Kitching et al. (2011) applied the LoVerde \& Afshordi (2008)
approximation (equation \ref{bl}) in the spherical-Bessel case and compared the 
case of full $(k,z)$ integration with the $\ell>kr$ case, 
and found a $< 10\%$ change in the amplitude of $C^{SB}_{\ell}(k,k')$ using the Limber approximation 
which was approximately constant as a function of $\ell$-mode, which is consistent with the results found in this paper.
Including the first and second order corrections suggested by LoVerde \& Afshordi (2008) are likely to 
reduce the impact further at low $\ell$-modes. 

Power spectrum methods, that measure the cosmic shear two-point statistics as a function of $\ell$-mode, are more immune 
to these approximations than correlation function methods 
because removing $\ell\ls 100$ from an analysis will eliminate most of the low-$\ell$ mode effects. 
This is the approach taken in K{\"o}hlinger et al. (2015) and Alsing et al. (2017) (both of which made the flat-sky, Limber 
and tomographic approximations). However power spectrum 
methods that use a pseudo-$C(l)$, or a mixing matrix method, to account for real-space masks will also encounter 
additional complexity if the masks mix low-$\ell$ modes and higher $\ell$-modes (e.g. Hikage et al., 2011). Finally 
super-sample covariance (Takada \& Hu, 2013) that causes correlations between the power 
spectrum errors across $\ell$-modes that will also mix low-$\ell$ and high-$\ell$ behaviour.  

\section{Conclusion}
\label{Conclusion}
In this paper we present the spherical-Bessel and spherical-radial representations of cosmic shear, and discuss the correlation 
function representation. We discuss several approximations and 
limits of these statistics including the flat-sky, tomographic and Limber approximations. 
Whilst the tomographic approximation is expected to be relatively benign -- because the lensing 
kernel is relatively smooth in redshift -- the flat-sky and Limber approximations change the statistical 
behaviour of the cosmic shear statistic at large-scales. 
We also find a subtlety in the derivation of the standard Limber-approximated cosmic shear power spectra formula 
that neglects an $\ell$-dependent factor of 
\ba
T_{\ell}=\frac{(\ell+2)(\ell+1)\ell(\ell-1)}{(\ell+0.5)^4}, 
\ea
which is equal to unity if the flat-sky approximation is used, and the factor of $0.5$ in the denominator is ignored. 
To include this effect any Limber-approximated cosmic shear potential power spectrum $C(\ell)$ 
should be multiplied by this factor (if not included already). 

We investigate how the angular scales in correlation function analyses map onto $\ell$-modes of the 
cosmic shear power spectrum and find that the following scaling relations are a good fit to the 
behaviour
\ba
\xi_+:\log_{10}[\ell_{\rm max}]&=&-0.14\log_{10}(\theta_{\rm min}/{\rm arcmin})+4.06\nn
\xi_-:\log_{10}[\ell_{\rm max}]&=&-0.19\log_{10}(\theta_{\rm min}/{\rm arcmin})+4.49.
\ea
We also present mapping between the various cosmic shear statistics used in the literature. In translating from the shear power spectrum to configuration statistics such as shear correlation functions, the Hankel transform introduces errors on arcminute scales and higher.  A full summation over spherical harmonic modes, using Wigner small-d matrices, is straightforward and preferable.

Many of the approximations we have discussed have relatively small effects, but are unnecessary and there is no good reason to apply them, and for future experiments, such as Euclid, LSST and 
WFIRST, which will have very small statistical errors, they should not be applied.  Only the Limber approximation 
may be necessary, and only if computational speed is an issue, and in this case the 
inaccuracies at low $\ell$ may be reduced by 
considering the first two terms in the Limber expansion in LoVerde \& Afshordi (2008).

In this paper we addressed the most 
prominent approximations, however there are several further approximations 
that are expected to have additional impacts on cosmological inference such as  
source-source clustering (Schneider et al. 2002), source-lens clustering
(Bernardeau 1998, Hamana et al. 2002), the Born approximation (Cooray \& Hu, 2002),
higher-order power spectrum terms (Krause \& Hirata, 2010), and the full treatment of unequal-time correlations (Kitching \& Heavens, 2016). 


\noindent{\em Acknowledgements:} TDK is supported by Royal Society University Research Fellowship. 
RJ \& LV acknowledge support by Spanish Mineco grant AYA2014-58747-P and MDM-2014-0369 of ICCUB (Unidad de 
Excelencia `Maria de Maeztu' and Royal Society grant IE140357. JDM is supported 
in part by the Engineering and Physical Sciences Research Council 
(grant number EP/M011852/1). The Centre for Computational Astrophysics is supported by the Simons Foundation. 
We thank the creators of {\tt CAMB} for public 
release of this code. We thank M. Cropper, H. Hoekstra, A. Lewis, and P. Paykari 
for useful and constructive discussions. We thank C. Wallis for providing the Wigner small-d matrices. 



\onecolumn

\section*{Appendix A: Shear Correlation Functions on the Sphere}
In this Appendix we derive equation (\ref{cof0}), that is the shear correlation on the celestial sphere. 
A spin-2 shear field may be written (see e.g. Hu 2000, Appendix A) 
\be
\label{a1}
\gamma_1(\hat n)\pm {\rm i}\gamma_2(\hat n)
=\frac{1}{2}\sum_{\ell m}[\phi^R_{\ell m}\pm{\rm i}\phi^I_{\ell m}]\sqrt{\frac{(\ell+2)!}{(\ell-2)!}} _{\pm 2}Y^m_{\ell}(\hat n)
\ee
where $\phi_{\ell m}$ is the spherical harmonic transform of the lensing potential with 
real and imaginary components, $_{\pm 2}Y^m_{\ell}(\hat n)$ are 
spin-2 spherical harmonics, and $\hat n$ are angular celestial coordinates. The shear
power spectrum is related to the lensing potential power spectrum by
\ba
C^{E,\gamma\gamma}_{\ell}&=&\frac{1}{4}\frac{(\ell+2)!}{(\ell-2)!}C^{\phi^R\phi^R}_{\ell}\nn
C^{B,\gamma\gamma}_{\ell}&=&\frac{1}{4}\frac{(\ell+2)!}{(\ell-2)!}C^{\phi^I\phi^I}_{\ell}
\ea
i.e. the E and B-mode are related to correlations of the real and imaginary parts of the lensing potential. 
To compute $\xi_+$, it is easiest to consider two points that are at the same azimuthal angle, separated by an angle 
in the polar direction. In this case 
$\xi_+=\langle(\gamma_1+{\rm i}\gamma_2)(\gamma_1+{\rm i}\gamma_2)^*\rangle=
\langle\gamma_1(\hat n)\gamma_1(\hat n')\rangle+\langle\gamma_2(\hat n)\gamma_2(\hat n')\rangle$ and 
$\xi_-=\langle(\gamma_1+{\rm i}\gamma_2)(\gamma_1-{\rm i}\gamma_2)^*\rangle=
\langle\gamma_1(\hat n)\gamma_1(\hat n')\rangle-\langle\gamma_2(\hat n)\gamma_2(\hat n')\rangle$, 
with $\hat n$ and $\hat n'$ separated by $\beta$, and
\ba
\xi_+(\beta)&=&\langle(\gamma_1+{\rm i}\gamma_2)(\gamma_1+{\rm i}\gamma_2)^*\rangle\nn
&=&\frac{1}{4}\sum_{\ell m\ell' m'}(\langle\phi^R_{\ell m}\phi^R_{\ell' m'}\rangle+\langle\phi^I_{\ell m}\phi^I_{\ell' m'}\rangle)
\frac{(\ell+2)!}{(\ell-2)!} {}_2Y^m_{\ell}(\hat n) {}_2Y^{m'*}_{\ell'}(\hat n')\nn
&=&\frac{1}{4}\sum_{\ell m}[C^{\phi^R\phi^R}_{\ell}+C^{\phi^I\phi^I}_{\ell}]\frac{(\ell+2)!}{(\ell-2)!}\sum_m {}_2Y^m_{\ell}(\hat n) {}_2Y^{m*}_{\ell}(\hat n')\nn
&=&\sum_{\ell}[C^{E,\gamma\gamma}_{\ell}+C^{B,\gamma\gamma}_{\ell}]\sqrt{\frac{2\ell+1}{4\pi}} {}_2Y^{-2}_{\ell}(\beta,0)
\ea
where the last inequality comes from Hu \& White (1997), equation (7) (with $\alpha=\tilde\gamma=0$; note that $\tilde\gamma$ here 
refers to an Euler angle not shear, but we use this as it is standard notation): 
\be 
\sum_m {}_2Y^m_{\ell}(\hat n) {}_2Y^{m*}_{\ell}(\hat n')=\sqrt{\frac{2\ell+1}{4\pi}}{}_2Y^{-2}_{\ell}(\beta,0). 
\ee
In terms of Wigner-D matrices, 
\be 
D^{\ell}_{\,\,\, -m s}(\alpha,\beta,-\tilde\gamma)=(-1)^m \sqrt{\frac{4\pi}{2\ell+1}}{}_sY^m_{\ell}(\beta,\alpha){\rm e}^{{\rm i}s\tilde\gamma}, 
\ee
hence
\be 
\xi_{+}(\beta)=\sum_{\ell}\left(\frac{2\ell+1}{4\pi}\right)[C^{E,\gamma\gamma}_{\ell}+C^{B,\gamma\gamma}_{\ell}]D^{\ell}_{22}(0,\beta,0), 
\ee
or in a more compact form in terms of small-d Wigner matrices 
\be
\xi_+(\beta)=\frac{1}{2\pi}\sum_{\ell}(\ell+0.5) d^{\ell}_{22}(\beta)[C^{E,\gamma\gamma}_{\ell}+C^{B,\gamma\gamma}_{\ell}]. 
\ee
A similar calculation for $\xi_-(\beta)$ is trivial by replacing the $+$ with $-$ in the derivation corresponding to the 
other case in equation (\ref{a1}). These results can also be derived trivially from Ng \& Liu (1999) equations (4.5-4.8) 
by identifying CMB polarisation quantities with their shear analogs. 

\section*{Appendix B: The Extended Limber Approximation for Cosmic Shear}
In LoVerde \& Afshordi (2008) an extended Limber approximation is presented that was used to assess the accuracy of this 
approximation as a function of $\ell$-mode. Their main result can be captured in the following approximation 
\be 
\label{la}
{\rm lim}_{\epsilon\rightarrow 0}\int_0^{\infty} {\rm e}^{-\epsilon(x-\nu)}f(x)J_{\nu}(x) {\rm d}x=
f(\nu)-\frac{1}{2}f''(\nu)-\frac{\nu}{6}f'''(\nu)+\dots
\ee
where $\nu=\ell+1/2$, $J_{\nu}(x)$ are Bessel functions (not spherical), and $f(x)$ is some arbitrary 
function. Dashes denote derivatives with respect to $x$. 
This is then applied to the case of a non-evolving matter power spectrum $P(k)$ 
(LoVerde \& Afshordi, 2008; equation 5) and an extended Limber approximation computed (LoVerde \& Afshordi, 2008; equation 11).
This calculation however is not strictly appropriate for the cosmic shear case because the matter power spectrum is 
an evolving field $P(k,z)$. 

For cosmic shear we start with equation (\ref{U})
\be
U_{\ell}(r[z],k)=\int_0^{r[z]} {\rm d}r' \frac{F_K(r,r')}{a(r')} j_{\ell}(kr')P^{1/2}(k,r'),
\ee
that describes the kernel function for the spherical-Bessel and spherical-radial representations of the cosmic shear 
field. The integral is along a line-of-sight to a source redshift plane $r[z]$ and encodes the radial 
transform of the integrated lensing 
effect caused by perturbations in the matter over-density, that are mapped to the power spectrum via Poissons equation. 
To make this into a form for which the LoVerde \& Afshordi (2008) expansion can be applied we re-write this as 
\be
U_{\ell}(r[z],k)=\int_0^{\infty} {\rm d}r' w(r[z],r') \frac{F_K(r,r')}{a(r')} j_{\ell}(kr')P^{1/2}(k,r'),
\ee
where $w(r,r')$ is a weight function with the following properties: $w(r,r')=1$ for $r'\leq r$, 
and $w(r,r')=0$ for $r'>r$. 
We can now apply the Limber approximation and find that to first order 
\be 
\label{ulim}
U^L_{\ell}(r[z],k)=\left(\frac{\pi}{2\nu k^2}\right)^{1/2} w(r[z],\nu/k) \frac{F_K(r,\nu/k)}{a(\nu/k)} P^{1/2}(k,\nu/k)+\dots
\ee
where $\nu=\ell+1/2$; and the pre-factor is a result of the conversion from a spherical Bessel function to a Bessel function. 
It can be seen explicitly that the weight function is now 
$w(r,\nu/k)=1$ for $\nu/k\leq r$, and $w(r,\nu/k)=0$ for $\nu/k>r$. 

The expansion of this case to higher orders can be done using equation (\ref{la}), however it can be seen that the 
calculation is more complex than the LoVerde \& Afshordi (2008) derivation because the function 
$f(x)$ in that equation is now $f(r'|r,k)=(\pi/2k^3r)^{1/2}w(r[z],r')[F_K(r,r')/a(r')]P^{1/2}(k,r')$. In particular the 
expansion does not affect the weight function evaluation 
$w(r,\nu/k)$ because the arguments to this do not change in the higher order terms.   
Also the expansion is only valid over the region $\nu/k< r$ where the derivatives of this function are not divergent. 

We can now attempt to derive the standard weak lensing formulation of the Limber-approximated cosmic shear power spectrum (Kaiser, 1998) using equations (\ref{sr}) and (\ref{Gsr}). Substituting equation (\ref{ulim}) we find that 
\ba
C^{SR,L}_{\ell}(z_i, z_j)= &&|D_{\ell}|^2{\mathcal A}^2\left(\frac{2}{\pi}\right)\int\frac{{\rm d}k}{k^2} \int {\rm d}z_p{\rm d}z'{\rm d}z'_p{\rm d}z'' 
n(z_p)n(z'_p)p(z'|z_p)p(z''|z'_p)W^{SR}(z_i,z_p)W^{SR}(z_j,z'_p)\nn
&&\left(\frac{\pi}{2\nu k^2}\right) \frac{F(r(z'),\nu/k)}{a(\nu/k)}\frac{F(r(z''),\nu/k)}{a(\nu/k)}P(k,\nu/k)
\ea
where we have absorbed the weight functions $w$ into the integral limits for clarity, and 
${\mathcal A} = 3\Omega_{\rm M} H_0^2/(2 c^2)$. 
To express this equation in the standard form we need to transform integration variables from $k$ to $r$ in the outer 
integral. This leads to 
\ba
C^{SR,L}_{\ell}(z_i, z_j)=&&|D_{\ell}|^2{\mathcal A}^2 \left(\frac{1}{\nu^4}\right)
\int {\rm d}r r^2 \int {\rm d}z_p{\rm d}z'{\rm d}z'_p{\rm d}z''
n(z_p)n(z'_p)p(z'|z_p)p(z''|z'_p)W^{SR}(z_i,z_p)W^{SR}(z_j,z'_p)\nn
&&\frac{F(r(z'),r)}{a(r)}\frac{F(r(z''),r)}{a(r)}P(\nu/r,r),
\ea
the inner integrals can now be expressed in terms of kernel functions 
\be
\label{qeq}
q(r_i,r)=\frac{r}{a(r)}\int {\rm d}z_p{\rm d}z'n(z_p)p(z'|z_p)W^{SR}(z_i,z_p)\left(\frac{r(z')-r}{r(z')}\right)
\ee
where we have expanded the function $F_K$ for the flat-geometry case ($K=0$), where 
the Limber-approximated power spectrum can be written as 
\be 
\label{at1}
C^{SR,L}_{\ell}(z_i, z_j)=|D_{\ell}|^2{\mathcal A}^2 \left(\frac{1}{\nu^4}\right)
\int {\rm d}r \frac{q(r_i,r)q(r_j,r)}{r^2} P(\nu/r,r).
\ee
This is the standard form for the cosmic shear power 
spectrum (see e.g. Hu, 1999; Joachimi \& Bridle, 2010), except that 
the $\ell$-dependent pre-factor is different. The full $\ell$-mode dependent prefactor is 
\ba 
\label{at2}
T_{\ell}=\frac{|D_{\ell}|^2}{\nu^4}=\frac{(\ell+2)(\ell+1)\ell(\ell-1)}{(\ell+0.5)^4}.
\ea
In the standard derivation there are two assumptions that remove this pre-factor. These assumptions are
the flat-sky approximation whereby $|D_{\ell}|^2=\ell^4$, and the approximation $\nu=\ell$ (or $\ell=(\ell+0.5)$). 
In this case $T_{\ell}=1$ and the standard result
is recovered. However these approximations can have a large impact on the amplitude of the 
power spectrum at $\ell\ls 100$ as we 
investigate in this paper. 
\begin{figure}
\centering
\includegraphics[angle=0,clip=,width=0.5\columnwidth]{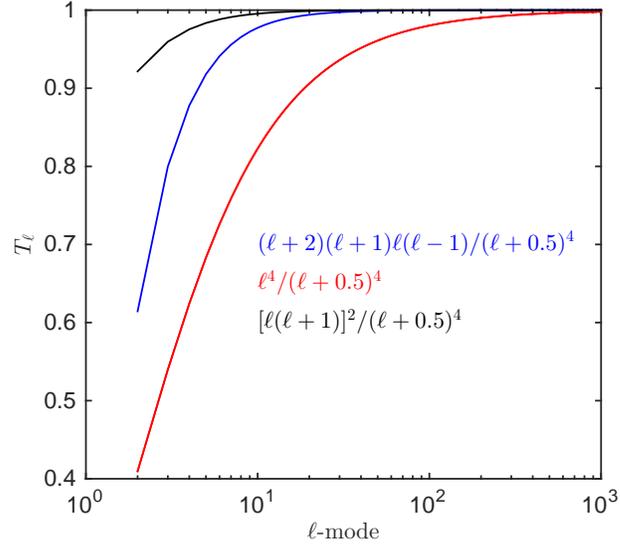}
\caption{The functional form of the $\ell$-dependent prefactor in equations (\ref{at1}), (\ref{at2}) and (\ref{at3}), for the 
cosmic shear spherical (blue) and flat-sky (red) cases, and for the convergence case (black).}
\label{tl_effect2}
\end{figure}
In Figure \ref{tl_effect2} we show the functional form of $T_{\ell}$. 
To recover the correct $\ell$-mode scaling from a standard cosmic shear analysis one should multiply by $T_{\ell}$. 

One can also compute a convergence power spectrum from weak lensing data. This is different from the shear 
case only in that the factor $D_{\ell}=\ell(\ell+1)$ in the spherical-sky case. 
Following the derivation above we find that the 
Limber-approximated convergence power spectrum is the same as equation (\ref{at1}) but with an $\ell$-dependent prefactor of 
\ba
\label{at3}
T^{\kappa}_{\ell}=\frac{[\ell(\ell+1)]^2}{(\ell+0.5)^4}.
\ea
Again, under the assumption that $\nu=\ell$ and $\ell\simeq \ell+1$ this factor cancels, but does not in general 
as also noted by Joudaki \& Kaplinghat (2012). 
We again show the effect in Figure \ref{tl_effect2}, which is less pronounced than for the shear case. 

\end{document}